\documentclass[preprint2,twoside]{hwo}

\usepackage{graphicx}
\usepackage{amsmath}
\usepackage{etoolbox}

\input{hwo.h}

\begin{document}

\title{\textbf{\LARGE Exploring Solar System Giant Planets with Habitable Worlds Observatory}}
\author {\textbf{\large Leigh N. Fletcher$^{1}$, Amy Simon$^2$, Michael H. Wong$^3$, Jonathan D. Nichols$^1$, Nick A. Teanby$^4$, Conor A. Nixon$^2$, Marina Galand$^5$}}
\affil{$^1$\small\it School of Physics and Astronomy, University of Leicester, University Road, Leicester, LE1 7RH, United Kingdom}
\affil{$^2$\small\it NASA Goddard Space Flight Center, 8800 Greenbelt Road, Greenbelt, MD 20771, USA.}
\affil{$^3$\small\it University of California, Berkeley, CA 94720, USA}
\affil{$^4$\small\it School of Earth Sciences, University of Bristol, Wills Memorial Building, Queens Road, Bristol, BS8 1RJ, UK.}
\affil{$^5$\small\it Department of Physics, Imperial College London, London SW7 2AZ, UK}

% Please add the names of endorsers in the format "Joseph Jensen (Utah Valley University), " separated by commas.
\author{\footnotesize{\bf Endorsed by:}
John Clarke,
Thierry Fouchet, 
Heidi B. Hammel,
Ricardo Hueso, 
Patrick G.J. Irwin, 
Caitriona Jackman,
Tommi Koskinen, 
Henrik Melin, 
Luke Moore, 
Julianne Moses
Imke de Pater, 
Glenn S. Orton, 
Larry Sromovsky,
Tom Stallard 
}

% This section is for ADS Processing.  There must be one line per author. Leave them commented out for the present. They will be included later.
%\paperauthor{Sample~Author1}{Author1Email@email.edu}{ORCID_Or_Blank}{Author1 Institution}{Author1 Department}{City}{State/Province}{Postal Code}{Country}
%\paperauthor{Sample~Author2}{Author2Email@email.edu}{ORCID_Or_Blank}{Author2 Institution}{Author2 Department}{City}{State/Province}{Postal Code}{Country}
%\paperauthor{Sample~Author3}{Author3Email@email.edu}{ORCID_Or_Blank}{Author3 Institution}{Author3 Department}{City}{State/Province}{Postal Code}{Country}

% Please provide entries for the Author index; leave them commented out for now.
%\aindex{Pacucci, F.}

\begin{abstract}
Visible and ultraviolet imaging and spectroscopy of Solar System giant planets can set the paradigm for the atmospheric, ionospheric, and magnetospheric processes shaping the diversity of giant exoplanets, brown dwarfs, and their interactions with stellar hosts.  Spectra of their molecular absorptions, aerosol scattering, airglow, and auroral emissions can reveal these dynamic atmospheres in three dimensions, from the cloud-forming weather layer, to the ionosphere and beyond.  Given that giant planets are extended, bright, moving, and rotating objects, with extreme dynamic range and highly variable appearances, they impose specific mission and instrumentation requirements on future large space-based optical/UV observatories like the proposed Habitable Worlds Observatory (HWO).  We advocate that HWO must have the capability to track non-sidereal targets like the giant planets and their satellites; should be able to view auroras and atmospheres without saturation (e.g., through the use of filters or fast read-out modes); and with a high dynamic range to explore faint objects near bright discs.  HWO should enable spatially-resolved spectroscopy from $\sim80$ nm to $\sim900$ nm, capturing H$_2$ Lyman and Werner band series and H Lyman-$\alpha$ in the far-UV; molecular absorptions and scattering in the mid-UV/visible; and deep hydrogen/methane absorptions in the 800-900 nm for cloud characterisation and CH$_4$ mapping.  Imaging should enable time-resolved observations, from seconds to create auroral movies, to hours for cloud tracking and winds, to months and years for atmosphere/ionosphere variability.   We advocate that an imager should have sufficient field of view to capture Jupiter ($>50$\arcsec), and that UV/visible integral field spectrographs be considered with both narrow ($3$\arcsec) and wide ($>10$\arcsec) field capabilities to provide efficient mapping of atmospheres and auroras.  A large field of regard would enable viewing giant planets over a full apparition, enabling rapid viewing of unexpected phenomena (impacts, storms, solar events, etc.), and maximal spatial resolution at opposition.  The benefits of extension beyond 1 $\mu$m are also discussed.  With these capabilities, HWO would enable new discoveries in comparative planetology, from our Solar System giants to those in other stellar systems.  \textit{This article is an adaptation of a science case document developed for HWO's Solar System Steering Committee.} \\
\end{abstract}

\vspace{2cm}

\section{Science Background and Goals}

The four Solar System Giant Planets represent the end-products of billions of years of planetary evolution, migration, and cooling, and are our closest and best examples of two categories of astrophysical objects known to be commonplace across our universe:  hydrogen-rich gas giants (Jupiter and Saturn), and heavy-element-rich ice giants (Uranus and Neptune).  

These worlds are planetary-scale laboratories, enabling exploration of the dynamics, circulation, and meteorology of objects from stars, to planets, and oceans.  Their deep chemical composition provides a window onto the distant past and the reservoirs of ices, gases, and dust available to the forming planets in the protoplanetary disc.  The composition and distribution of gases and aerosols in their visible atmospheres, from the cloud-forming 'weather layers' (troposphere), to the photochemical factories of their middle atmospheres (stratosphere), reveal the chemical and seasonal processes shaping their environments.  And emissions from their upper atmospheres provide a critical window for processes at work in their wider systems (satellites, moons, and magnetospheres), revealed via auroras and airglow, to understand how each world interacts with both their satellites and the solar wind.  Finally, their ever-changing atmospheres capture the imagination of professionals and public alike, making them prime targets for the transformative new capabilities of any future observatory.

\begin{figure*}[ht!]
\centering
\includegraphics[width=\textwidth]{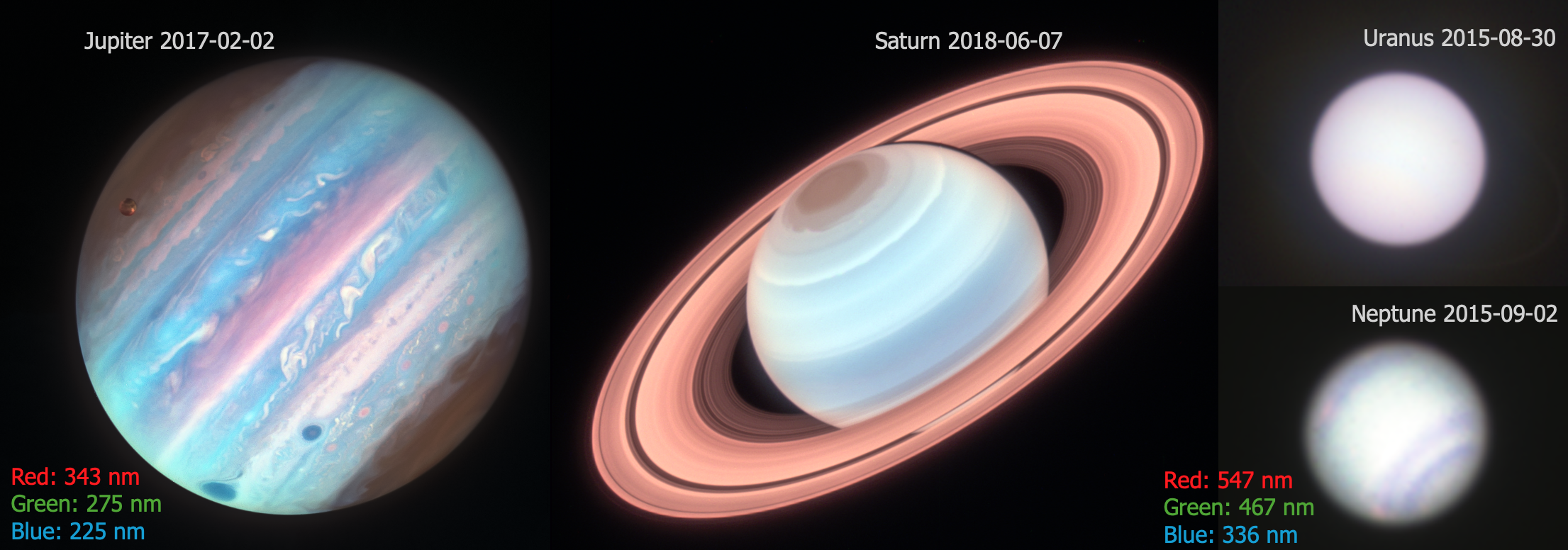}
\caption{The four giant planets as seen by Hubble WFC3/UVIS at different near-UV wavelengths, and processed by Judy Schmidt (note - these images are not shown to scale).  The Jupiter and Saturn images are constructed from F343N (red), F275W (green), and F225 (blue), whereas a lack of observations for Uranus and Neptune required us to shift to redder wavelengths:  F547M (red), F467M (green), and F336W (blue) for the Ice Giants.  Jupiter and Ganymede were observed 2017-02-02 at 12:16UT, Saturn was observed at 2018-06-07 at 00:06UT, Uranus on 2015-08-30 at 10:08UT, and Neptune on 2015-09-02 at 07:58UT.  Strong UV absorption is visible at the poles of Jupiter and Saturn due to the nature of high-altitude aerosols and the potential link to auroral chemistry.  Red vortices, such as Jupiter's Great Red Spot and Oval BA (lower left), appear dark due to the blue-absorbing chromophore responsible for their red colours.  Uranus' bright north polar cap of aerosols, and its narrow rings, are just visible in this image.  Neptune's banding, and some bright high clouds (appearing in pink) can be seen in both northern and southern hemispheres.  In addition, Saturn's rings appear reddish due to UV absorption.  Credit: NASA/ESA/CSA/Hubble/Judy Schmidt. }
\label{fig_HubbleUV}
\end{figure*}

\begin{figure*}[ht!]
\centering
\includegraphics[width=0.9\textwidth]{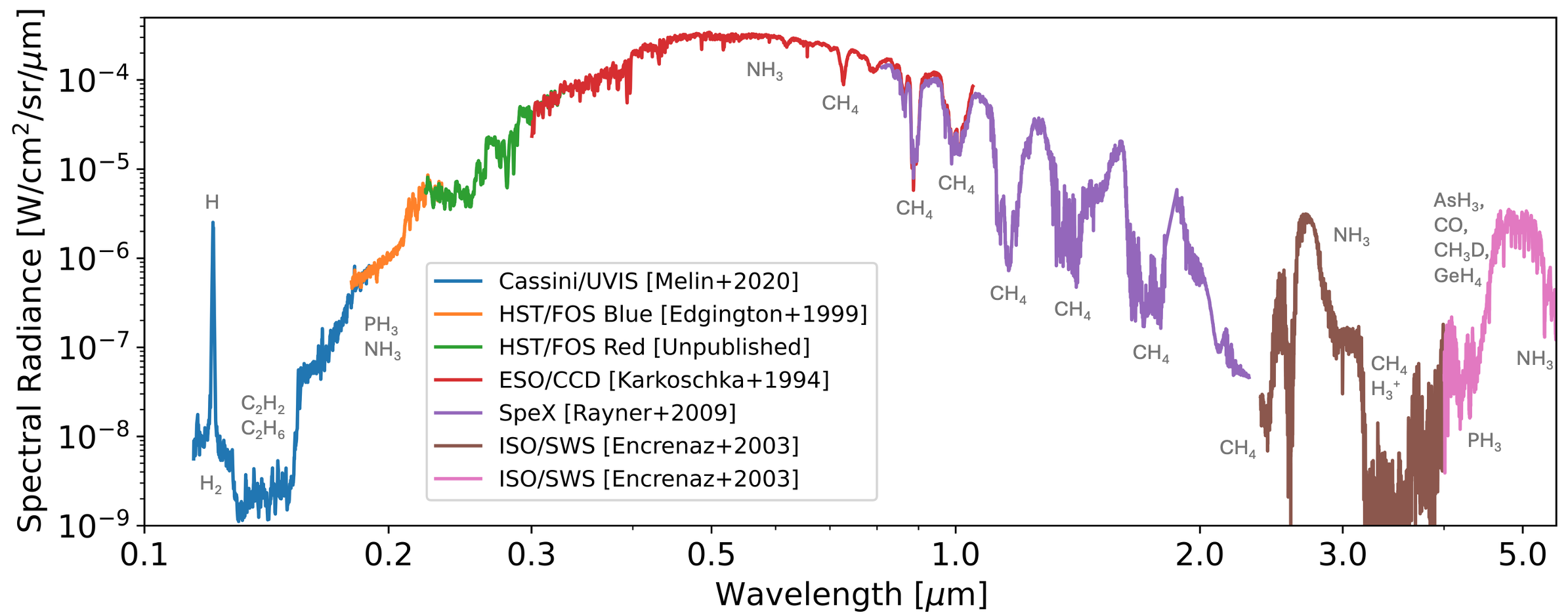}
\caption{Montage of Jupiter spectra from the UV \citep{20melin_uv, 99edgington}, visible \citep{94karkoschka}, and near-infrared \citep{09rayner, 03encrenaz}, adapted from \citet{23fletcher}. Key atmospheric absorptions and emissions have been labelled.  We define the spectral ranges as follows: extreme UV (50-120 nm); far-UV (121-200 nm), mid-UV (200-300 nm), near-UV (300-400 nm), visible ($>400$ nm), near-IR ($>900$ nm), and the start of the mid-IR at $\sim5$ $\mu$m. }
\label{fig_spectra}
\end{figure*}

Giant planet UV and visible spectra are shaped by reflected sunlight, scattered (via Rayleigh, Raman, and Mie scattering) by the aerosols and gases in their tropospheres and stratospheres, and energetic particle precipitation \citep{04taylor, 04moses, 04clarke, 09fouchet, 18fletcher_book, 22simon, 23sanchez-lavega}.  Examples of the UV appearance of the four giant planets, from 225-547 nm, are shown in Hubble imaging in Fig. \ref{fig_HubbleUV}, with representative UV, visible, and IR spectra shown for Jupiter in Fig. \ref{fig_spectra}.  The reflected sunlight spectrum displays absorptions due to a host of gaseous species - cloud-forming volatiles like ammonia (NH$_3$) and methane (CH$_4$), hydrogen (H$_2$), disequilibrium species mixed upwards from the deeper interior like phosphine (PH$_3$), and the hydrocarbon products of carbon photochemistry in the stratosphere, like ethane (C$_2$H$_6$) and acetylene (C$_2$H$_2$).  These absorption features sound different altitudes, enabling the tracking of meteorological activity (clouds, winds, plumes, and vortices) at different depths within the atmospheres.  The slope of the UV-visible spectra enable studies of the chromophores responsible for colouring the clouds and hazes, along with the sizes and distributions of those aerosols.  At shorter wavelengths, emissions from H$_2$ Lyman and Werner bands (80-180 nm), hydrogen Lyman-$\alpha$ (121.6 nm), as well as helium (50-60 nm), enable exploration of neutral atmospheres via airglow, and of planetary auroras and the energetic processes that drive them \citep{79sandel, 80clarke, 01grodent, 09clarke, 17bonfond, 17nichols, 17gladstone, 24benmahi, 25chaufray_hwo}.  Furthermore, Cassini and Hubble investigations of Saturn have shown that H Ly-$\alpha$ reveals seasonal changes to the atomic hydrogen population in the upper atmosphere, reflecting changes to both photochemistry and dynamics \citep{23benjaffel}.  On Uranus, the H Ly-$\alpha$ also tracks Rayleigh scattering by H$_2$ in the middle atmosphere \citep{25joshi}, in addition to the upper-atmospheric hydrogen population.  Synergistic observations from these various wavelengths therefore reveals the vertical coupling between the different layers of the atmosphere.

For all these reasons, ultraviolet (UV), visible, and infrared (IR) spectroscopy of planetary atmospheres and ionospheres has been a key component of spacecraft payloads designed to visit each world.  Voyagers 1 and 2 carried the Ultraviolet Spectrometer to all four planets (UVS, 50-170 nm, capturing auroral, airglow, and occultations of UV-bright stars), along with the imaging camera (ISS, with filters between 280-640 nm).  Galileo carried the Ultraviolet Spectrometer (UVS, 113-432 nm), Extreme Ultraviolet Spectrometer (EUV, 54-128 nm), and Solid State Imager (SSI, eight filters spanning 375-1100 nm) to Jupiter between 1995 and 2003, followed by Juno's UVS (68-210 nm) and JunoCam (filters in the RGB and strong CH$_4$ band at 889 nm) since 2016. ESA's Jupiter Icy Moons Explorer (JUICE) and NASA's Europa Clipper will also have capabilities spanning the UV (both UVS instruments span 55-210 nm, for both spectral mapping and solar/stellar occultations) and visible imaging (JUICE/JANUS covering 380-1080 nm, Clipper/EIS covering $\sim$360-1000 nm).  At Saturn, the Cassini spacecraft had an Ultraviolet Imaging Spectrograph (UVIS, 56-190 nm) and the Imaging Science Subsystem (ISS, 200-1100 nm), with some of this range also covered by the visible and infrared mapping spectrometer (VIMS, 250-1000 nm in the visible channel).

These close-in observations have been supplemented by UV and visible data from orbital observatories \citep{22simon}, including the International Ultraviolet Explorer between 1978 and 1996 \citep[IUE, 115-320 nm,][]{80owen, 85wagener}, and the range of instrumentation on Hubble.  The Faint Object Spectrograph (FOS, 115-850 nm) was used to explore the distribution of tropospheric and stratospheric species on Jupiter and Saturn \citep{99edgington}.  FOS was replaced by STIS (Space Telescope Imaging Spectrograph, 115-1000 nm) in 1997,  capable of both imaging and long-slit spectroscopy, and used extensively to study auroras on Jupiter \citep[e.g.,][]{16gustin} and Saturn \citep{13gerard}, as well as the clouds and composition of Uranus \citep{11sromovsky} and Neptune \citep{11karkoschka_ch4}.  Since 2014, all four giant planets have been observed annually as part of the Outer Planet Atmospheres Legacy programme \citep[OPAL,][]{15simon}, using Hubble's Wide Field Camera 3 (WFC3).  Example images of the four giant planets, viewed between 225-547 nm by WFC3/UVIS, are shown in Fig. \ref{fig_HubbleUV}.

In this short article, we will not attempt to review the giant planet science from these spacecraft missions and observatories.  However, we have used them as a guide for the spacecraft and instrumentation requirements that could be considered for the Habitable Worlds Observatory, to ensure that this cutting-edge facility of the 2040s is capable of observing our four giant planets.  Our scientific goals can be summarised as follow:

\begin{itemize}

\item Explore the dynamical, chemical, and cloud-forming processes shaping the atmospheres and ionospheres of the Solar System gas giants (Jupiter, Saturn) and ice giants (Uranus, Neptune) as the archetypes for this class of planetary object.  
\item Explore how energy is transported vertically and horizontally, and from the smallest to the largest scales, by winds, waves, magnetosphere-ionosphere coupling, and large-scale circulation patterns from the troposphere to the thermosphere.
\item Explore how the planets are coupled to magnetospheric and solar wind plasmas, influencing atmospheric properties (auroral emission, heating, chemistry, electrical conductivity, dynamics, and aerosols).

\end{itemize}

While some of these goals will be familiar to those working on Jupiter and Saturn, the capabilities of HWO will be transformative for our exploration of Uranus and Neptune, potentially long before these worlds get their first dedicated orbital spacecraft. HWO will accelerate our exploration of the Ice Giants, from their atmospheres to their auroras, enabling true comparative planetology with their larger Gas Giant cousins. The Ice Giant Systems (i.e., including their `ocean world' satellite systems and rings) are a top priority for the US Decadal Survey in Planetary Science and Astrobiology; and the interaction between planetary magnetospheres and solar/stellar winds remains a key priority for the US Heliophysics Decadal Survey.  HWO exploration of Solar System Giants therefore provides the potential for discoveries at the interface of space science disciplines.  

In the following sections, we summarise the objectives of ultraviolet, visible, and infrared observations of giant planet atmospheres and ionospheres (Section \ref{obj}), describe the physical parameters that HWO should try to measure (Section \ref{params}), and finish with a list of HWO observatory and instrument requirements that should be met to enable discovery-level science for the gas and ice giants in the decades to come (Section \ref{reqs}).  For the purpose of this article, we assume that HWO will have a 6-8 m primary mirror, with wide-field imaging from the far UV ($\sim100$ nm) to the near-IR (2.2 $\mu$m), and UV-visible spectroscopy (preferably via an integral field spectrometer, IFS) from 80-900 nm.  The extension to 900 nm is necessary to distinguish aerosols and methane variability on Ice Giants, and to sound aerosols as a function of height from the troposphere into the lower stratosphere.

\section{Science Objective}
\label{obj}

In this section, we provide a high-level overview of the objectives of ultraviolet, visible, and short-wave infrared imaging and spectroscopy for giant planet atmospheres.  The physical parameters required to address these objectives, and the resulting requirements on the observatory and instrumentation, are provided in the following sections.

\begin{figure*}[ht!]
\centering
\includegraphics[width=0.75\textwidth]{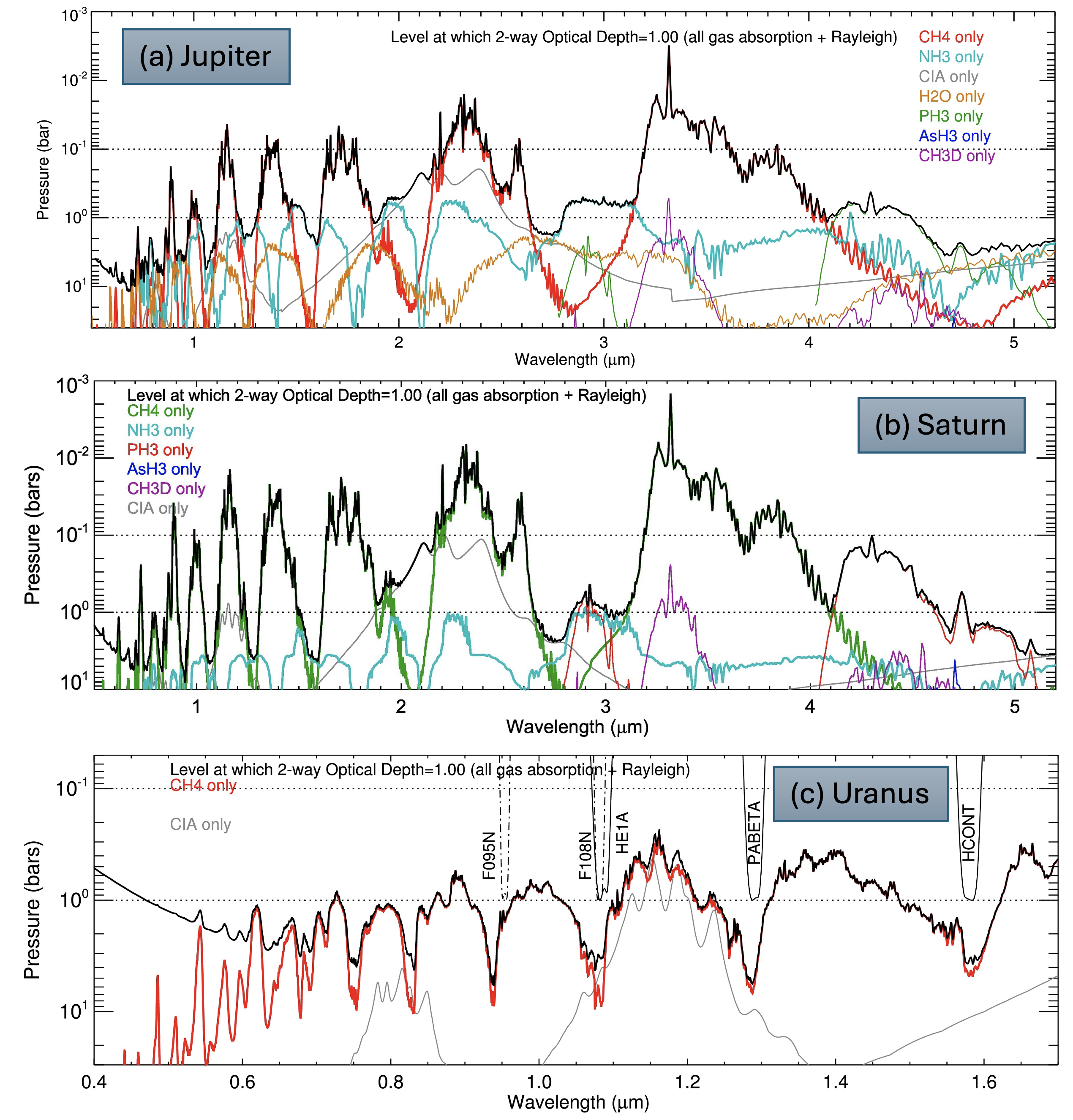}
\caption{Two-way transmission functions, indicating the approximate penetration depth to an optical depth of unity in an aerosol-free atmosphere, reproduced with permission for Jupiter \citep{18sromovsky_jup}, Saturn \citep{20sromovsky} and Uranus \citep{19sromovsky}.   The full 0.4-5.2 $\mu$m range is shown for Jupiter and Saturn, but we restrict to 0.4-1.7 $\mu$m for Uranus (which also indicates the location of filters used by Hubble and Keck to measure the distribution of CH$_4$).  Multiple gases are shown for Jupiter and Saturn, only CH$_4$ and the H$_2$, He continuum are shown for Uranus.  Rayleigh scattering dominates at shorter wavelengths.}
\label{fig_transmission}
\end{figure*}

\subsection{Imaging Objectives}

Global imaging of the giant planets in narrow-band filters from the UV to the near-IR enables an exploration of dynamics through cloud-tracking, resolving motions in the zonal (east-west) and meridional (north-south) direction, constructing vorticity fields (a diagnostic of fluid motion) and kinetic energy spectra (to explore how energy cascades from the scale of the smallest eddies up to the largest jet streams and vortices, and vice versa).  This investigation requires narrow-band imaging in filters sensing in and out of strong methane bands in the visible and near-IR, as shown in Fig. \ref{fig_transmission}. Continuum regions in relatively transparent windows can sense condensate clouds of methane and hydrogen sulphide  on the Ice Giants, and ammonia, NH$_4$SH and potentially water-ice clouds on the Gas Giants.  Conversely, imaging in strong methane bands provides access to cloud and haze features in the upper troposphere and stratosphere, as recently exploited by NIRCam on JWST \citep{23hueso}, as well as filters sensing UV absorption/scattering for the highest altitudes.  Cloud tracking requires repeated views with a separation of 1-2 hours to observe the motions of features.  

\begin{figure*}[ht!]
\centering
\includegraphics[width=\textwidth]{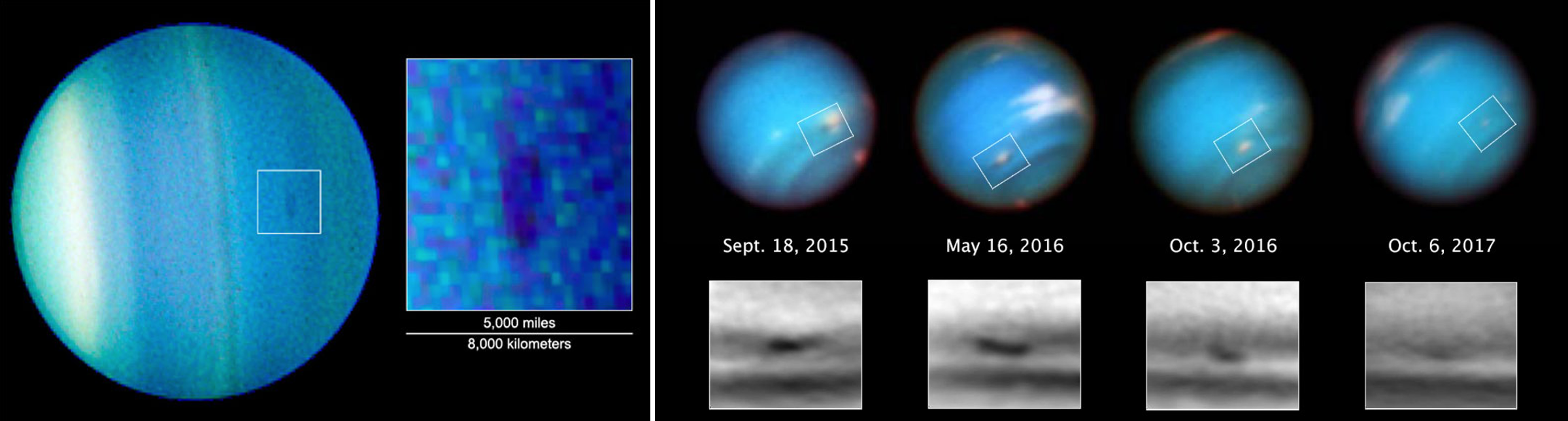}
\caption{Dark ovals on Uranus and Neptune, which are best explored with HWO's exceptional spatial resolution at blue wavelengths.  Uranus and its dark spot \citep[left,][]{09hammel} was observed on 2006-08-23 in three filters - red 775 nm, green 658 nm, blue 550 nm [Credit NASA, ESA, L. Sromovsky and P. Fry (University of Wisconsin), H. Hammel (Space Science Institute), and K. Rages (SETI Institute)].  The shrinkage of Neptune's southern dark spot (SDS-2015, not to be confused with the Great Dark Spot observed by Voyager) is shown in three-colour images top right (red 763 nm, green 547 nm, blue 467 nm), and in just the blue filter bottom right to reveal the dark oval \citep{18wong} [Credit: NASA, ESA, and M.H. Wong and A.I. Hsu (UC Berkeley)].}
\label{fig_ovals}
\end{figure*}

Some of the giant planet science described in this paper can only be accomplished through UV and `blue' imaging.  For example, auroral emissions and airglow on giant planets can only be explored in the far-UV, and dark vortices on the Ice Giants have their highest contrast at blue (400-500 nm) wavelengths, as shown in Fig. \ref{fig_ovals}.  Large dark ovals exhibit dramatically different behaviour on the Ice Giants compared to Jupiter's Great Red Spot, with their shorter lifetimes (years, rather than centuries) dynamically linked to their latitudinal drift over time.  Very little is known about their thermal, chemical, and aerosol properties, how they form and dissipate, and how they compare to their gas-giant counterparts.  Several examples have now been tracked in Hubble imaging on Neptune \citep{18wong}, and spectra were obtained from VLT/MUSE \citep{23irwin}, but only one example of a dark oval has been seen on Uranus \cite{09hammel}.  HWO imaging at blue wavelengths will be exceptional for the exploration of these unique Ice Giant dynamical features.

In addition, imaging will provide:
\begin{itemize}
\item Global views of contrasts across the discs, repeated on a variety of timescales to explore time-variable processes (storms, plumes, and planetary impact scars evolving over hours; changes in the banded structure evolving over months; seasonally dependent aerosol changes over years).

\item Auroral images with narrow-band filters focused on H$_2$ Lyman and Werner bands (80-180 nm), and H Lyman-$\alpha$, watching evolution over short timescales (seconds to explore auroral morphology) and medium/long-term timescales (days to ~years to establish relations between auroral brightness, solar wind conditions and plasma production at Io, through solar minimum/maximum).

\item High-sensitivity and high-resolution views of impact flashes/bolides \citep{18hueso_impact} (if impactors are detected in advance by facilities such as the Vera Rubin Observatory) and their resulting UV-dark scars \citep{10hammel}.  Although impact events have been explored previously on Jupiter by amateur and professional facilities like Hubble, they have not yet been detected on Saturn or the Ice Giants. UV and visible imaging of the impact debris serves as an excellent tracer of atmospheric dynamics in regions high above the cloud-tops, at altitudes where wind tracking is usually hampered by a lack of discrete clouds.  The extreme sensitivity of HWO could allow monitoring of much smaller impact events than previously observed, resulting in new estimates of the external comet/meteoroid flux at the giant planets. 

\end{itemize}

\subsection{Spectroscopy Objectives}
  
Although narrow-band UV and visible imaging provides excellent insights into atmospheric dynamics, the true transformative impact of HWO for giant planets will be via spatially resolved mapping spectroscopy, with the aim of providing global coverage (i.e., full coverage of the visible hemisphere as the planet rotates) with repetition tuned to a variety of timescales (e.g., as above for imaging).  Observations from the disc-centre to the edge will provide constraints on the scattering/absorption properties of aerosols in the UV-visible, as a function of latitude (clouds and hazes vary with latitude due to different dependence on condensation, temperature, auroral influence, etc.).  

Spectroscopy of the giant planets enables the following:
\begin{itemize}

\item Aerosol characterisation:  Determining the vertical distribution of aerosols as a function of disc location, constraining their wavelength dependent scattering/absorption properties; particle size distributions/shapes; and possibly their composition via comparisons to refractive index data measured in laboratories.

\item Tropospheric composition:  mapping of species such as ammonia, phosphine, and methane provides insights on dynamics (e.g., the belt/zone structure and inter-hemispheric circulations), condensation (e.g., volatile species forming cloud decks, like ammonia on Jupiter, or methane on Neptune), and disequilibrium chemistry (e.g., mixing of phosphine from the deeper troposphere, and photochemical depletion due to UV irradiation).  HWO spectroscopy enables measurements of the vertical profiles of these species \citep{99edgington}, and their relation with the clouds and hazes of the troposphere.

\item Stratospheric composition: Mapping of hydrocarbons (ethane, acetylene, propane, diacetylene, and others with UV-visible cross-sections) provides access to the rich chemical factories of their stratospheres.  Comparative planetology then reveals the influence of insolation, obliquity, seasons, and the strength of vertical diffusion on the distribution of hydrocarbons from planet to planet.  Correlation of hydrocarbons with auroral energy input reveals the influence of ion chemistry on stratospheric composition, and potentially on ion influx at lower latitudes (e.g., `ring rain' on Saturn).  HWO spectroscopy will provide access to both horizontal distributions and vertical profiles for some stratospheric species \citep{06prange}, which will aid in understanding vertical transport processes to constrain chemical modelling. 

\item Chemistry and dynamics from cometary and asteroidal impacts:  As mentioned above, impact events have dramatic consequences for planetary atmospheres.  The SL9 impact on Jupiter generated a rich exotic cocktail of species from high temperature shock-chemistry of the impact \citep{04harrington}. There is also compelling evidence for a cometary impact on Neptune from trace species that could only generated in a large impact \citep{17moreno}. HWO spectroscopy would allow detailed study of these species and their temporal evolution, thus revealing unique information about giant planet upper atmosphere chemical processes and lifetimes.

\item Auroral energy and temperature:  As described by \citet{25chaufray_hwo}, UV-visible spectroscopy of planetary aurora reveals the altitudes, extent, and variability of auroral energy deposition into the atmospheres.  Medium-resolution (e.g., 0.01 nm at 120 nm, $R\sim12,000$) would provide hydrogen temperatures, precipitating electron energies and energy flux, and would enable measurements of spectral line profiles to determine upper atmospheric winds.  Doppler shifting of auroral emission features (i.e., for ion/neutral winds) would require ultra-high UV spectral resolution (for H$_2$ moving at 0.1 km/s, we would need $R>10^5$), so this is likely out of reach of HWO unless an additional slit mode can be envisioned (a detailed tradeoff study would be required to assess this), or as a component of a multi-object spectrograph.  

\item Stellar occultations: Although the number of FUV-bright stars is limited, the tracking of UV-bright stars as they are occulted by the planetary disc (being attenuated by delicate haze layers on the planetary limbs) may enable measurement of the temperature and density of the middle and upper atmosphere, as well as the vertical distributions of chemical species if the star is bright enough.  Stellar occultations have been used to good effect from orbiters such as Cassini/UVIS for Saturn \citep{20brown}, and will be a key observing strategy for ESA's JUICE mission at Jupiter \citep{23fletcher}. Cooler stars can also be used to probe temperature profiles via refraction, but will not provide chemical distributions.  The availability of bright stars for HWO occultations should be studied in future work.

\item Planetary energy balance: The bond albedo is a key parameter for estimating the net balance of absorbed solar heating versus emitted power, revealing the self-luminosity of all four giant planets due to their slow cooling over time.  The wavelength range of HWO covers the peak in the reflected solar component, at wavelengths inaccessible to Voyager.  In particular, solar contributions in the near-UV were neglected in some previous estimates of the energy balance, leading to significant uncertainty in the size of the internal heat sources.  This is important for all the giant planets, but especially for Uranus, where the internal heat source is small and the emitted and absorbed fluxes are only slightly out of balance \citep{25irwin_ura}.  HWO will be able to significantly refine the magnitude of this energy imbalance, which is critical for understanding interior heat transport processes and internal structure.

\end{itemize}

\section{Physical Parameters and Proposed Measurements}
\label{params}

The objectives above require the following measurements, broken down into imaging and spectroscopy measurements.  UV-Visible imaging requirements are as follows:
\begin{itemize}
\item Imaging of cloud motions in multiple filters sounding from deep tropospheric clouds to haze layers in the stratosphere.  Rapid repeat imaging (over timescales of hours) is needed to determine meridional and zonal velocities, and their relationship with the banded structure and storms/vortices.   Over longer timescales, regular imaging sequences (over years) are needed to observe how the vorticity fields, kinetic energy, and spatial scales of features all vary with time.
\item Multi-wavelength imaging of atmospheric features to diagnose their dynamics and aerosol properties, targeting global views (e.g., requiring wide fields of view) and discrete features (cyclones and anticyclones, wave features, storm plumes, impact scars, polar domains, etc.).  UV mapping accesses the absorptive/scattering properties of high-altitude aerosols; visible imaging accesses the opacity and colours of the deeper clouds.
\item Multi-wavelength auroral imaging with short timescales:  Movies of planetary auroras to diagnose the rich dynamics of the auroras and their coupling to the wider planetary system (e.g., mapping to processes in the magnetosphere, coupling to plasma tubes from satellites, and influence of variable solar wind conditions).  Provided the imager is capable of observing in the FUV, then HWO will be capable of UV imaging of planetary auroras surpassing that possible from Hubble.  
\item Stellar occultations:  Monitoring as UV-bright stars are occulted by the planetary limbs, taking high frequency measurements as the upper atmospheric layers refract the starlight, enabling derivations of temperature/density profiles at high spatial resolution.
\end{itemize}

UV-Visible mapping spectroscopy requirements are as follows:
\begin{itemize}
\item Globally map the spatial distribution of UV and visible-absorbing species in the tropospheres (ammonia, phosphine, methane) and stratospheres (hydrocarbons), allowing derivations of aerosol properties (vertical distribution and layering, wavelength-dependent refractive indices, potentially aerosol composition) and abundance profiles of chemical species.
\item Spectroscopy of auroral emission: determine the auroral electron energy characteristics and induced atmospheric heating, and how they change spatially and temporally.
\item Stellar occultations:  Repeat the occultation experiment with UV-bright stars to provide high-resolution vertical constraints on temperature, density, aerosols and possibly chemical composition through the occultation points.
\item Repeat all spectroscopic measurements over regular time intervals (months to years) to catch the evolution of atmospheric phenomena and explore both seasonal and non-seasonal processes.
\end{itemize}

As giant planet atmospheres evolve over a range of timescales, we propose a series of HWO campaigns tuned to specific phenomena.  Short-term campaigns could provide auroral imaging to understand how they respond to conditions in their magnetospheres and the solar wind, and to intercompare the substantially different auroral processes on gas and ice giants (the latter having asymmetric, offset, and multipolar magnetic fields; and with each system displaying different couplings between the planets and their satellite systems).  Repeated observations over months and years could observe how the winds and chemical distributions change in response to meteorology and seasonal variations.  An extreme solar event could have impacts on several planets (if aligned), so that HWO could observe as the event propagates from one planet to the next \citep{04prange}.  And, following the successful example of Hubble/OPAL, repetition over years and decades would reveal how the seasonal giants (Saturn, Uranus, and Neptune) respond to changes in insolation, or the changing orientation of their magnetic axes with respect to the Sun, over their long orbits.

%\subsection{Key Physical Parameters}

%\subsection{Observational State of the Art and Science Targets}

\section{Measurement Requirements for Habitable Worlds Observatory}
\label{reqs}

The UV-visible imaging and spectroscopy proposed above requires an observatory, and its suite of instruments, to be designed with observations of extended, rotating, bright, moving targets in mind.  Although we focus on capabilities in the UV and visible, we briefly also mention the benefits of extension into the near-infrared, overlapping with the capabilities of JWST's NIRSpec instrument ($3\times3$\arcsec field of view and a spectral resolution of $R\sim2700$ from 0.97 to 5.3 $\mu$m).

\begin{figure*}[ht!]
\centering
\includegraphics[width=0.75\textwidth]{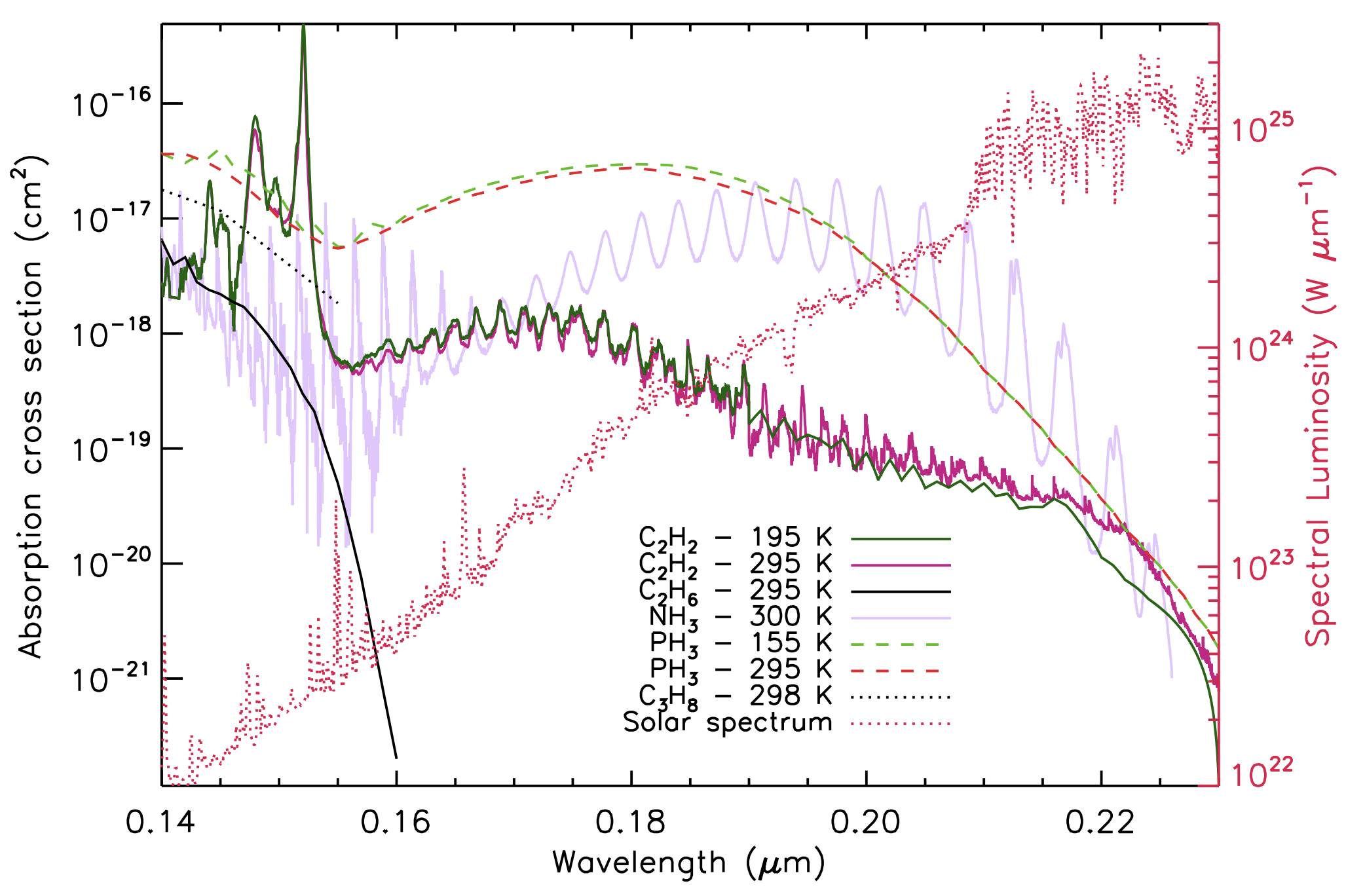}
\caption{Absorption cross-sections for a number of important chemical species in giant planet atmospheres, compared to the solar luminosity (red, dashed line), reproduced from \citet{20melin_uv}.  Laboratory measurements at different temperatures are shown for acetylene (C$_2$H$_2$) and phosphine (PH$_3$), with single measurements for ethane (C$_2$H$_6$), ammonia (NH$_3$), and propane (C$_3$H$_8$).  Methane (CH$_4$), carbon monoxide (CO), diacetylene (C$_4$H$_2$), and water (H$_2$O) also have UV absorptions in this region.}
\label{fig_UVxsection}
\end{figure*}

\begin{itemize}

\item \textbf{Moving Target tracking and mosaics:} HWO must have the ability to guide and track non-sidereal rotating/moving giant planets to enable `stare' observations and mosaicking, as well as the ability to track moving satellites and background stars as they are occulted by the planetary disc. HWO must be able to track specific surface features on rotating objects, such as specific storms, auroral ovals, or impact scars. Moving target tracking and mosaicking have been essential for Solar System science from Hubble and JWST, and if HWO is able to track small asteroids, near-Earth objects, comets and interstellar objects near perihelion, then it should be able to track objects within the giant planet systems.

\item \textbf{Bright Target Capability:} HWO should have the ability to view Jupiter (and other giants) at all wavelengths without saturation.  JWST saturates on Jupiter at many IR wavelengths, and even on Neptune for some of the brightest cloud features, making analysis challenging.  Depending on the sensitivity of HWO's instruments, this might require consideration of neutral density filters for imaging and spectroscopy, and/or the use of very short read-out times for HWO detectors, and photon-counting capability.

\item \textbf{Frequency of Observation and Repetition:} HWO should be capable of performing repeat imaging over 10-20 hrs (20-40 hrs for improved precision) for cloud tracking and auroral variability.  The accuracy of tracking cloud/eddy motion increases with multiple measurements. This is especially needed on the Ice Giants where wind fields are poorly known due to the small numbers of cloud features observed in the visible.

\item \textbf{Long-term monitoring:}	HWO should enable repeat observations of planetary atmospheres and ionospheres to create a  long-term legacy, following the successful example of Hubble/OPAL.  This could entail annual views of the giant planets with imaging and spectroscopy to track slowly evolving climates and relation to solar wind conditions.  Long-term monitoring is key - these are not static worlds, but evolve over seconds, hours, months, years, and decades.

\item \textbf{UV spectroscopy:} Spectroscopy and imaging should span EUV (50-120 nm); FUV (121-200 nm, shown in Fig. \ref{fig_UVxsection}), MUV (200-300 nm), NUV (300-400 nm) and visible ($>400$ nm). Extension into EUV ($<120$ nm) and FUV gives access to H$_2$ Lyman series (80 to 180 nm), Werner band systems and H Lyman-$\alpha$, hydrocarbons, and aerosols.  Then MUV ($>160$ nm) gives PH$_3$, NH$_3$ and Raman absorptions.  The NUV enables studies of chromophores and aerosol characteristics, dark ovals on Ice Giants, as well as CH$_4$ absorptions for cloud tomography. Furthermore, the EUV extension to 50-60 nm provides access to Helium emission lines.

\item \textbf{UV time-resolved imaging capability for auroras:}  High-speed imaging of variable auroras requires imaging in the far-UV with a suitable field of view to capture the auroral region (e.g., $>20$\arcsec).  For example, the detector used by Hubble for FUV auroral observations is the STIS/MAMA detector, a solar-blind CsI MAMA, with $\sim0.025$ arcsecond ­pixels, and a nominal $25\times25$ arcsecond square FOV, operating in the FUV from 115 to 170 nm (Jupiter observers typically use the 25MAMA clear filter and F25SRF2 long-pass filter which has a low wavelength cut-off at 127.5 nm).

\item \textbf{Visible Observations (imaging and spectroscopy):}	 Narrow-band imaging and spectroscopy should span the visible (400-900 nm) range for CH$_4$, NH$_3$, and H$_2$ absorptions. Filters for the HWO imager should consider (and be tuned to) the UV/Vis absorption spectra of giant planets. For example, we should ensure that filters sound the following features at a minimum:  the strongest CH$_4$ bands near 890 nm for high-altitude clouds; access to 825-nm H$_2$ features to separate CH$_4$ and clouds on ice giants; 600-700 nm NH$_3$ features on Jupiter/Saturn.  Observations in the visible enable colour discrimination, determination of aerosol distributions (in and out of CH$_4$ bands), and mapping of chemical distributions (CH$_4$, NH$_3$).

\item \textbf{Infrared observations ($> 1.0$ microns):} Measurements of aerosols and gaseous composition would significantly benefit from the addition of a near-IR capability to HWO.  At a minimum, this should reach out to 1.6 $\mu$m to sample CH$_4$, H$_2$S, and H$_2$ absorption features. Extension to 3.0 $\mu$m would capture the prominent 2.7-$\mu$m reflectivity peak (sounding deep clouds) and evidence for ice absorptions. This extension to the infrared would enable mapping of volatiles (NH$_3$, H$_2$S, CH$_4$) on the ice giants to reveal their general circulation and planetary banding.  Extension to 3.5 $\mu$m would enable mapping of auroral H$^{+}_{3}$ emission.  Simultaneous auroral observations in FUV and NIR are a powerful combination to address the science described above, being able to observe both energy inputs (UV) and outputs (IR) at the same time.  Although there is clear overlap with NIRCam and NIRSpec on JWST, this proposed infrared extension would continue the legacy of JWST, and perhaps avoid the saturation issues that have presented a challenge for Solar System observations.

\item \textbf{Spatial Resolution (imaging and spectroscopy):}	To provide significant advances over existing capabilities, we wish to sample to smallest cloud/eddy/auroral scales possible in the UV and visible. This requires imaging at better than 50 km/px on Jupiter to surpass Cassini winds/turbulence movies \citep{03porco}.  For spectroscopy, the current state of the art is with the MUSE instrument on the VLT, which has been used to view Jupiter and Saturn with $0.2\times0.2$\arcsec spaxels sampling a $60\times60$\arcsec field of view \citep{25irwin}.  Thus HWO integral field spectroscopy in the UV and visible could consider wide-field modes (over a $>10$\arcsec field with a 0.2\arcsec spaxel scale) and narrow-field modes (0.02\arcsec per spaxel over a 3\arcsec field) in order to advance the state of the art.  The PSF should be very well characterised to enable extraction of features at the diffraction limit of the telescope (assuming an 8-m primary, the spatial resolution of 0.02\arcsec at 650 nm would provide 60, 120, 255 and 430 km spatial resolution at Jupiter, Saturn, Uranus and Neptune oppositions, respectively). 

\item \textbf{Spectral Resolution and Range:} At visible wavelengths, HWO spectroscopy should aim to surpass MUSE ($R\sim1740$ at 480 nm, $R\sim3450$ at 930 nm), and in the ultraviolet we require $R>3000$ for spectral emission and absorption features.  It would be desirable to match or exceed HST/STIS ($R>10^4$) and MUSE to enable measurement of line widths (and hence winds), whilst ensuring that the instrument can sample as broad a spectral range as possible. Higher spectral resolutions enable greater discrimination of temperatures and electron energies of auroral emissions (H$_2$, He) and atmospheric absorptions (PH$_3$, NH$_3$, CH$_4$, H$_2$).  Atmospheric retrievals need to have a broad band coverage, so the HWO instrument should avoid very narrow spectral ranges that require multiple settings to provide a complete spectrum.  Furthermore, electron energies can be measured if the spectral range is wide enough, as it is determined by the ratio of intensities near 160 and 125 nm.

\item \textbf{Field of View (FOV, Imaging):}  Given the large angular diameters of the giant planets as seen from Earth, special consideration of the FOVs of instruments is required to prevent the need for burdensome mosaicking.  Global imaging of Jupiter's atmosphere requires a FOV of 50\arcsec, but 120\arcsec could be used to capture the wider jovian system (rings, satellites, plasma torii, etc.).  The Ice Giants are more straightforward, with angular sizes of 3.7\arcsec and 2.2\arcsec for Uranus and Neptune, respectively, but their wider systems (rings, moons, etc.) would also benefit from FOVs exceeding 10\arcsec.  Like JWST/NIRCam, subarrays could be used to avoid saturation if imaging with a smaller FOV is possible.

\item \textbf{Field of View (UV-Visible spectroscopy, assuming IFS):}	 Although capturing global views of the giant planets may be unfeasible, experience has shown that larger FOVs enable better location of specific clouds and meteorological features, with moderate mosaicking used to map planets with JWST in the infrared.  We propose an IFS FOV no smaller than 3\arcsec (with spaxel scales smaller than 0.05\arcsec) to enable mapping of Uranus and Neptune.  A larger field of 10 to 20\arcsec (with spaxel scales smaller than 0.3\arcsec) would enable easier mosaicking of Jupiter and Saturn, capture Jupiter's entire auroral region at once, and enable observation of Uranus with its rings.  A compromise with two IFS modes, one narrow and one wide, would provide context for regional mapping.

\item \textbf{Large Field of Regard (FoR):}  At a minimum, HWO must be able to view the four giant planets at some point along their orbits, with observations near opposition enabling the maximal spatial resolution for tracking small cloud features, particularly on the Ice Giants.  JWST observations are restricted to quadrature, due to the nature of the sunshield, meaning there are only two FoR windows for each planet per year, lasting 50 days.  This is unsuitable for reacting to new and unexpected phenomena, such as impact events, or storm eruptions.  Ideally, HWO would have the capability to observe giant planets at any point during their apparitions (e.g., from quadrature to quadrature).	

\item \textbf{Rapid Response:}	Hubble, JWST, and ground-based facilities enable target-of-opportunity observations of new and unexpected phenomena, such as impact events and new storm systems, usually within approximately 7 days, but faster responses are preferred.  These are high-science-value events, but with a low probability of occurrence within a single year, so HWO should instigate `community triggers' from the outset:  fully-designed observations with zero proprietary time, outside of the usual telescope allocation process, so that HWO can be ready for the unexpected. 

\item \textbf{Dwell time and sampling frequency:}  HWO should be capable of staring at a planet for multiple hours to generate `movies,' using sub-second frame times to observe the variability of planetary auroras and evolving cloud features over many hours.  This ability to focus on a single target for hours has not been possible with Hubble due to the nature of its orbit, nor with spacecraft flybys due to their short-term observations.  HWO could allow continuous observations for much longer timeframes.

\item \textbf{High dynamic range:}. Giant planet systems will challenge HWO due to the large variety of surfaces reflecting sunlight.  The observatory instruments need to have sufficient dynamic range to be able to view dark objects next to very bright ones without saturation or scattered light (e.g., dark vortices in bright bands, or fine haze layers on the planetary limb).  JWST diffraction spikes and scattered light create a significant challenge for Solar System observers. For JWST, planetary observers have had the ability to rotate the position angle of the observatory to isolate and remove any diffraction spike effects. A key test is the ability to observe small inner moons next to the glare of the main planets; and Saturn at methane-band wavelengths next to bright rings.  	

\end{itemize}

\section{The Need for HWO}

Given that the four giant planets are readily visible in our night sky, and can be observed by ground-based telescopes from amateurs and professional alike, it is fair to ask why the premier facility of the 2040s will be needed to advance giant planet science.  The key advances are expected to be in the UV, which will `go dark' after the eventual demise of Hubble.  Spatially-resolved UV spectroscopy, coupled with UV-visible imaging from a substantially larger primary mirror, will enable new capabilities and discoveries, particularly for the Ice Giants, Uranus and Neptune.  Indeed, the Ice Giant atmospheres, ionospheres, rings, and moons will be observed at a unprecedented spatial resolution by HWO, and possibly long before the arrival of dedicated robotic orbiters for these distant worlds.  By exploring the dynamics, chemistry and energetics of their atmospheres; the interaction with the solar wind and magnetosphere via auroras; the surface chemistry and geomorphology of their icy moons; and the gravitational interactions and composition of their planetary rings, HWO will address questions spanning US decadal surveys in planetary science, heliophysics, and astrophysics, along with Europe's Voyage-2050 strategic roadmap, among others.  We strongly encourage the HWO team to consider capabilities for long-term observations of these bright, rotating, moving, extended targets from the outset, to enable new giant planet discoveries in the decades to come.

Finally, although we did not discuss satellites or rings in this short paper, there is a host of other intriguing phenomena that can be also explored with HWO, ranging from potential habitability (icy satellites), extreme volcanism (Io), to the effects of plasma irradiation and solar radiation on surfaces and atmospheres.  After Hubble and JWST,  HWO will undoubtedly be the next telescope to bring about transformational science, not only for the atmospheres and ionospheres discussed here, but for the entirety of these complex and ever-changing giant planet systems.

%\acknowledgements
{\bf Acknowledgements.}  
Fletcher was supported by STFC Grant reference UKRI/ST/B001205/1, Nichols was supported by STFC grant reference UKRI1206.  We are grateful to Judy Schmidt for providing the processed UV images of the giant planets in Fig. \ref{fig_HubbleUV}, and to Larry Sromovsky for providing the two-way transmission functions in Fig. \ref{fig_transmission}. For the purpose of open access, the author has applied a Creative Commons Attribution (CC BY) licence to the Author Accepted Manuscript version arising from this submission. 

\bibliography{references.bib}

\end{document}